\input epsf
\input amssym 


\newfam\scrfam
\batchmode\font\tenscr=rsfs10 \errorstopmode
\ifx\tenscr\nullfont
        \message{rsfs script font not available. Replacing with calligraphic.}
        \def\scr{\cal}
\else   
        \font\sevenscr=rsfs7
        \font\fivescr=rsfs5
        \skewchar\tenscr='177 \skewchar\sevenscr='177 \skewchar\fivescr='177
        \textfont\scrfam=\tenscr \scriptfont\scrfam=\sevenscr
        \scriptscriptfont\scrfam=\fivescr
        \def\scr{\fam\scrfam}
        \def\cal{\scr}
\fi
\catcode`\@=11
\newfam\frakfam
\batchmode\font\tenfrak=eufm10 \errorstopmode
\ifx\tenfrak\nullfont
        \message{eufm font not available. Replacing with italic.}
        
\else
    
    \font\sevenfrak=eufm7 \font\fivefrak=eufm5
    \textfont\frakfam=\tenfrak
    \scriptfont\frakfam=\sevenfrak \scriptscriptfont\frakfam=\fivefrak
    
\fi
\catcode`\@=\active
\newfam\msbfam
\batchmode\font\twelvemsb=msbm10 scaled\magstep1 \errorstopmode
\ifx\twelvemsb\nullfont\def\Bbb{\bf}

    \message{Blackboard bold not available. Replacing with boldface.}
\else   \catcode`\@=11
        \font\tenmsb=msbm10 \font\sevenmsb=msbm7 \font\fivemsb=msbm5
        \textfont\msbfam=\tenmsb
        \scriptfont\msbfam=\sevenmsb \scriptscriptfont\msbfam=\fivemsb
        \def\Bbb{\relax\expandafter\Bbb@}
        \def\Bbb@#1{{\Bbb@@{#1}}}
        \def\Bbb@@#1{\fam\msbfam\relax#1}
        \catcode`\@=\active

\fi
\newfam\cpfam
\def\sectionfonts{\relax
    \textfont0=\twelvecp          \scriptfont0=\ninecp
      \scriptscriptfont0=\sixrm
    \textfont1=\twelvei           \scriptfont1=\ninei
      \scriptscriptfont1=\sixi
    \textfont2=\twelvesy           \scriptfont2=\ninesy
      \scriptscriptfont2=\sixsy
    \textfont3=\twelveex          \scriptfont3=\tenex
      \scriptscriptfont3=\tenex
    \textfont\itfam=\twelveit     \scriptfont\itfam=\nineit
    \textfont\slfam=\twelvesl     \scriptfont\slfam=\ninesl
    \textfont\bffam=\twelvebf     \scriptfont\bffam=\ninebf
      \scriptscriptfont\bffam=\sixbf
    \textfont\ttfam=\twelvett
    \textfont\cpfam=\twelvecp
}
        \font\eightrm=cmr8              \def\xrm{\eightrm}
        \font\eightbf=cmbx8             \def\xbf{\eightbf}
        \font\eightit=cmti10 at 8pt     \def\xit{\eightit}
                       
        \font\sixrm=cmr6                
                     
        \font\eightcp=cmcsc8
        \font\eighti=cmmi8              \def\xold{\eighti}
        \font\eightib=cmmib8             \def\xbold{\eightib}
        \font\teni=cmmi10               \def\old{\teni}
        \font\ninei=cmmi9
        \font\tencp=cmcsc10
        \font\ninecp=cmcsc9

        \font\twelvei=cmmi12
        \font\twelvecp=cmcsc10 scaled\magstep1

        \font\twelvesy=cmsy12
        \font\ninesy=cmsy9
        \font\sixsy=cmsy6
        \font\twelveex=cmex12

        \font\twelveit=cmti12
        \font\nineit=cmti9
        
        \font\twelvesl=cmsl12
        \font\ninesl=cmsl9
        
        \font\twelvebf=cmbx12
        \font\ninebf=cmbx9
        \font\sixbf=cmbx6
        \font\twelvett=cmtt12

        \font\sixi=cmmi6

\batchmode\font\tenhelvbold=phvb at10pt \errorstopmode
\ifx\tenhelvbold\nullfont
        \message{phvb font not available. Replacing with cmr.}
    \font\tenhelvbold=cmb10   
    \font\twelvehelvbold=cmb12
    
    \font\sixteenhelvbold=cmb16
  \else
    \font\tenhelvbold=phvb at10pt   
    \font\twelvehelvbold=phvb at12pt
     at14pt
    \font\sixteenhelvbold=phvb at16pt
\fi

\def\noblackbox{\overfullrule=0pt}
\noblackbox

\newtoks\headtext
\headline={\ifnum\pageno=1\hfill\else
    \ifodd\pageno{\eightcp\the\headtext}{ }\dotfill{ }{\old\folio}
    \else{\old\folio}{ }\dotfill{ }{\eightcp\the\headtext}\fi
    \fi}
\def\makeheadline{\vbox to 0pt{\vss\noindent\the\headline\break
\hbox to\hsize{\hfill}}
        \vskip2\baselineskip}
\newcount\infootnote
\infootnote=0
\def\foot#1#2{\infootnote=1
\footnote{${}^{#1}$}{\vtop{\baselineskip=.75\baselineskip
\advance\hsize by
-\parindent\noindent{\xrm #2\hfill\vskip\parskip}}}\infootnote=0$\,$}
\newcount\refcount
\refcount=1
\newwrite\refwrite
\def\oldsize{\ifnum\infootnote=1\xold\else\old\fi}
\def\ref#1#2{
    \def#1{{{\oldsize\the\refcount}}\ifnum\the\refcount=1\immediate\openout\refwrite=\jobname.refs\fi\immediate\write\refwrite{\item{[{\xold\the\refcount}]}
    #2\hfill\par\vskip-2pt}\xdef#1{{\noexpand\oldsize\the\refcount}}\global\advance\refcount by 1}
    }
\def\refout{\catcode`\@=11
        \xrm\immediate\closeout\refwrite
        \vskip2\baselineskip
        {\noindent\twelvecp References}\hfill
        \par\nobreak\vskip\baselineskip
        \baselineskip=.75\baselineskip
        \input\jobname.refs
        \baselineskip=4\baselineskip \divide\baselineskip by 3
        \catcode`\@=\active\rm}

\def\hepth#1{\href{http://arxiv.org/abs/hep-th/#1}{arXiv:hep-th/{\xold#1}}}

\def\arxiv#1#2{\href{http://arxiv.org/abs/#1.#2}{arXiv:{\xold#1}.{\xold#2}}}
\def\jhep#1#2#3#4{\href{http://jhep.sissa.it/stdsearch?paper=#2\%28#3\%29#4}{J. High Energy Phys. {\xbold #1#2} ({\xold#3}) {\xold#4}}}

\def\CQG#1#2#3{Class. Quantum Grav. {\xbold#1} ({\xold#2}) {\xold#3}}

\def\JHEP{\jhep}

\def\NPB#1#2#3{Nucl. Phys. {\xbf B}{\xbold#1} ({\xold#2}) {\xold#3}}

\def\PLB#1#2#3{Phys. Lett. {\xbf B}{\xbold#1} ({\xold#2}) {\xold#3}}

\def\PRD#1#2#3{Phys. Rev. {\xbf D}{\xbold#1} ({\xold#2}) {\xold#3}}

\newcount\sectioncount
\sectioncount=0
\def\section#1#2{\global\eqcount=0
    \global\subsectioncount=0
        \global\advance\sectioncount by 1
    \ifnum\sectioncount>1
            \vskip2\baselineskip
    \fi
    \noindent
       \line{\sectionfonts\twelvecp\the\sectioncount. #2\hfill}
        \par\nobreak\vskip.8\baselineskip\noindent
        \xdef#1{{\old\the\sectioncount}}}
\newcount\subsectioncount
\def\subsection#1#2{\global\advance\subsectioncount by 1
    \par\nobreak\vskip.8\baselineskip\noindent
    \line{\tencp\the\sectioncount.\the\subsectioncount. #2\hfill}
    \vskip.5\baselineskip\noindent
    \xdef#1{{\old\the\sectioncount}.{\old\the\subsectioncount}}}
\newcount\appendixcount
\appendixcount=0
\def\appendix#1{\global\eqcount=0
        \global\advance\appendixcount by 1
        \vskip2\baselineskip\noindent
        \ifnum\the\appendixcount=1
        \hbox{\twelvecp Appendix A: #1\hfill}
        \par\nobreak\vskip\baselineskip\noindent\fi
    \ifnum\the\appendixcount=2
        \hbox{\twelvecp Appendix B: #1\hfill}
        \par\nobreak\vskip\baselineskip\noindent\fi
    \ifnum\the\appendixcount=3
        \hbox{\twelvecp Appendix C: #1\hfill}
        \par\nobreak\vskip\baselineskip\noindent\fi}
\def\acknowledgements{\vskip2\baselineskip\noindent
        \underbar{\it Acknowledgements:}\ }
\newcount\eqcount
\eqcount=0
\def\Eqn#1{\global\advance\eqcount by 1
\ifnum\the\sectioncount=0
    \xdef#1{{\old\the\eqcount}}
    \eqno({\oldstyle\the\eqcount})
\else
        \ifnum\the\appendixcount=0
            \xdef#1{{\old\the\sectioncount}.{\old\the\eqcount}}
                \eqno({\oldstyle\the\sectioncount}.{\oldstyle\the\eqcount})\fi
        \ifnum\the\appendixcount=1
            \xdef#1{{\oldstyle A}.{\old\the\eqcount}}
                \eqno({\oldstyle A}.{\oldstyle\the\eqcount})\fi
        \ifnum\the\appendixcount=2
            \xdef#1{{\oldstyle B}.{\old\the\eqcount}}
                \eqno({\oldstyle B}.{\oldstyle\the\eqcount})\fi
        \ifnum\the\appendixcount=3
            \xdef#1{{\oldstyle C}.{\old\the\eqcount}}
                \eqno({\oldstyle C}.{\oldstyle\the\eqcount})\fi
\fi}
\def\eqn{\global\advance\eqcount by 1
\ifnum\the\sectioncount=0
    \eqno({\oldstyle\the\eqcount})
\else
        \ifnum\the\appendixcount=0
                \eqno({\oldstyle\the\sectioncount}.{\oldstyle\the\eqcount})\fi
        \ifnum\the\appendixcount=1
                \eqno({\oldstyle A}.{\oldstyle\the\eqcount})\fi
        \ifnum\the\appendixcount=2
                \eqno({\oldstyle B}.{\oldstyle\the\eqcount})\fi
        \ifnum\the\appendixcount=3
                \eqno({\oldstyle C}.{\oldstyle\the\eqcount})\fi
\fi}
\def\multi{\global\advance\eqcount by 1}
\def\multieq#1#2{
    \ifnum\the\sectioncount=0
        \eqno({\oldstyle\the\eqcount})
         \xdef#1{{\old\the\eqcount#2}}
    \else
        \xdef#1{{\old\the\sectioncount}.{\old\the\eqcount}#2}
        \eqno{({\oldstyle\the\sectioncount}.{\oldstyle\the\eqcount}#2)}
    \fi}

\newtoks\url
\def\Href#1#2{\catcode`\#=12\url={#1}\catcode`\#=\active#2}
\def\href#1#2{{#2}}
\def\hhref#1{{#1}}
\parskip=3.5pt plus .3pt minus .3pt
\baselineskip=14pt plus .1pt minus .05pt
\lineskip=.5pt plus .05pt minus .05pt
\lineskiplimit=.5pt
\abovedisplayskip=18pt plus 4pt minus 2pt
\belowdisplayskip=\abovedisplayskip
\hsize=14cm
\vsize=20.2cm
\hoffset=1.5cm
\voffset=1.6cm
\frenchspacing
\footline={}
\raggedbottom

\def\ss{\scriptstyle}

\def\*{\partial}
\def\punkt{\,\,.}
\def\komma{\,\,,}

\def\={\!=\!}
\def\small#1{{\hbox{$#1$}}}

\def\fraction#1{\small{1\over#1}}
\def\fr{\fraction}
\def\Fraction#1#2{\small{#1\over#2}}
\def\Fr{\Fraction}

\def\eg{{\tenit e.g.}}

\def\ie{{\tenit i.e.}}

\def\a{\alpha}
\def\b{\beta}

\def\d{\delta}
\def\e{\varepsilon}
\def\g{\gamma}

\def\R{{\Bbb R}}




\def\l{\lambda}

\def\lra{\longrightarrow}

\def\arrowunder#1{\raise4pt\vtop{\baselineskip=0pt\lineskip=0pt
      \ialign{\hfill##\hfill\cr${\ss #1}$\cr$\lra$\cr}}}

\def\adj{\hbox{\bf adj}}
\def\xadj{\hbox{\sixbf adj}}
\def\R{\hbox{\bf R}}
\def\xR{\hbox{\sixbf R}}

\def\leftbr{[\hskip-1.5pt[}
\def\rightbr{]\hskip-1.5pt]}

\def\dslash{\partial\hskip-5pt/}

\def\<{{<}}
\def\>{{>}}


\ref\BaggerLambertI{J. Bagger and N. Lambert, {\xit ``Modeling
multiple M2's''}, \PRD{75}{2007}{045020} [\hepth{0611108}].}

\ref\BaggerLambertII{J. Bagger and N. Lambert, {\xit ``Gauge symmetry
and supersymmetry of multiple M2-branes''}, \PRD{77}{2008}{065008}
[\arxiv{0711}{0955}].} 

\ref\BaggerLambertIII{J. Bagger and N. Lambert, {\xit ``Comments on
multiple M2-branes''}, \JHEP{08}{02}{2008}{105} [\arxiv{0712}{3738}].}

\ref\Gustavsson{A. Gustavsson, {\xit ``Algebraic structures on
parallel M2-branes''}, \arxiv{0709}{1260}.}

\ref\Papadopoulos{G. Papadopoulos, {\xit ``M2-branes, 3-Lie algebras
and Pl\"ucker relations''}, \jhep{08}{05}{2008}{054}
[\arxiv{0804}{2662}].}

\ref\GauntlettGutowski{J.P. Gauntlett and J.B. Gutowski, {\xit
``Constraining maximally supersymmetric membrane actions''},
\hfill\break\arxiv{0804}{3078}.} 

\ref\LambertTong{N. Lambert and D. Tong, {\xit ``Membranes on an
orbifold''}, \arxiv{0804}{1114}.}

\ref\DMPvR{J. Distler, S. Mukhi, C. Papageorgakis and M. van
Raamsdonk, {\xit ``M2-branes on M-folds''}, \jhep{08}{05}{2008}{038}
[\arxiv{0804}{1256}].}

\ref\StringTalks{Talks by N. Lambert, J. Maldacena and S. Mukhi at
Strings 2008, CERN, Gen\`eve, August 2008, \hhref{www.cern.ch/strings2008}.}

\ref\CederwallNilssonTsimpisI{M. Cederwall, B.E.W. Nilsson and D. Tsimpis,
{\xit ``The structure of maximally supersymmetric super-Yang--Mills
theory---constraining higher order corrections''},
\jhep{01}{06}{2001}{034} 
[\hepth{0102009}].}

\ref\CederwallNilssonTsimpisII{M. Cederwall, B.E.W. Nilsson and D. Tsimpis,
{\xit ``D=10 super-Yang--Mills at $\ss O(\a'^2)$''},
\JHEP{01}{07}{2001}{042} [\hepth{0104236}].}

\ref\BerkovitsParticle{N. Berkovits, {\xit ``Covariant quantization of
the superparticle using pure spinors''}, \jhep{01}{09}{2001}{016}
[\hepth{0105050}].}

\ref\SpinorialCohomology{M. Cederwall, B.E.W. Nilsson and D. Tsimpis,
{\xit ``Spinorial cohomology and maximally supersymmetric theories''},
\jhep{02}{02}{2002}{009} [\hepth{0110069}];
M. Cederwall, {\xit ``Superspace methods in string theory, supergravity and gauge theory''}, Lectures at the XXXVII Winter School in Theoretical Physics ``New Developments in Fundamental Interactions Theories'',  Karpacz, Poland,  Feb. 6-15, 2001, \hepth{0105176}.}

\ref\Movshev{M. Movshev and A. Schwarz, {\xit ``On maximally
supersymmetric Yang--Mills theories''}, \NPB{681}{2004}{324}
[\hepth{0311132}].}

\ref\BerkovitsI{N. Berkovits,
{\xit ``Super-Poincar\'e covariant quantization of the superstring''},
\jhep{00}{04}{2000}{018} [\hepth{0001035}].}

\ref\BerkovitsNonMinimal{N. Berkovits,
{\xit ``Pure spinor formalism as an N=2 topological string''},
\jhep{05}{10}{2005}{089} [\hepth{0509120}].}

\ref\CederwallNilssonSix{M. Cederwall and B.E.W. Nilsson, {\xit ``Pure
spinors and D=6 super-Yang--Mills''}, \arxiv{0801}{1428}.}

\ref\CGNN{M. Cederwall, U. Gran, M. Nielsen and B.E.W. Nilsson,
{\xit ``Manifestly supersymmetric M-theory''},
\JHEP{00}{10}{2000}{041} [\hepth{0007035}];
{\xit ``Generalised 11-dimensional supergravity''}, \hepth{0010042}.
}

\ref\CGNT{M. Cederwall, U. Gran, B.E.W. Nilsson and D. Tsimpis,
{\xit ``Supersymmetric corrections to eleven-dimen\-sional supergravity''},
\jhep{05}{05}{2005}{052} [\hepth{0409107}].}

\ref\HoweTsimpis{P.S. Howe and D. Tsimpis, {\xit ``On higher order
corrections in M theory''}, \jhep{03}{09}{2003}{038} [\hepth{0305129}].}

\ref\NilssonPure{B.E.W.~Nilsson,
{\xit ``Pure spinors as auxiliary fields in the ten-dimensional
supersymmetric Yang--Mills theory''},
\CQG3{1986}{{\xrm L}41}.}

\ref\HowePureI{P.S. Howe, {\xit ``Pure spinor lines in superspace and
ten-dimensional supersymmetric theories''}, \PLB{258}{1991}{141}.}

\ref\HowePureII{P.S. Howe, {\xit ``Pure spinors, function superspaces
and supergravity theories in ten and eleven dimensions''},
\PLB{273}{1991}{90}.} 

\ref\FreGrassi{P. Fr\'e and P.A. Grassi, {\xit ``Pure spinor formalism
for OSp(N$\ss |$4) backgrounds''}, \arxiv{0807}{0044}.}


\headtext={M. Cederwall: ``N=8 superfield formulation of the BLG model''}

\line{
\epsfxsize=18mm
\epsffile{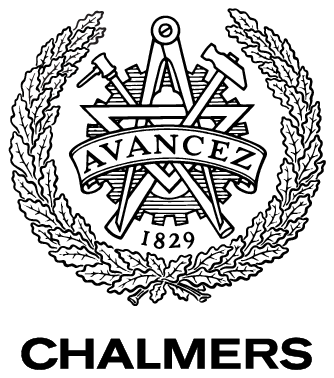}
\hfill}
\vskip-12mm
\line{\hfill G\"oteborg preprint}
\line{\hfill August, {\old2008}}
\line{\hrulefill}

\vfill
\vskip.5cm

\centerline{\sixteenhelvbold
N=8 superfield formulation} 

\vskip4\parskip

\centerline{\sixteenhelvbold
of the Bagger--Lambert--Gustavsson model}

\vfill

\centerline{\twelvehelvbold
Martin Cederwall}

\vfill

\centerline{\it Fundamental Physics}
\centerline{\it Chalmers University of Technology}
\centerline{\it SE 412 96 G\"oteborg, Sweden}

\vfill

{\narrower\noindent \underbar{Abstract:} We reformulate the
Bagger--Lambert--Gustavsson model using an $N=8$ superspace, thus
making the full supersymmetry manifest. The formulation is based on
appropriate ``pure spinor wave functions'' for the Chern--Simons and
matter multiplets. The Lagrangian has an extremely simple structure,
essentially containing a Chern--Simons-like term for the gauge field
wave function and a minimally coupled matter wave function. No higher
order interactions than cubic are present. The
consistency of the setup relies on an interplay between the algebraic
structures of the 3-algebra and of the pure spinors.
\smallskip}
\vfill

\font\xxtt=cmtt6

\vtop{\baselineskip=.6\baselineskip\xxtt
\line{\hrulefill}
\catcode`\@=11
\line{email: martin.cederwall@chalmers.se\hfill}
\catcode`\@=\active
}

\eject

\noindent The Bagger--Lambert--Gustavsson (BLG) model 
[\BaggerLambertI,\Gustavsson,\BaggerLambertII,\BaggerLambertIII] is
a maximally ($N=8$) supersymmetric and conformal interacting
3-dimensional model, whose local degrees of freedom consist of scalar
multiplets. It has been proposed to be related to AdS${}_4$ boundary
theories describing multiple membrane configurations, 
although the interpretation is unclear, partly due to the fact that
there only is one unique finite-dimensional representation for the
fields (an $so(4)$ gauge algebra with matter in the vector
representation) [\Papadopoulos,\GauntlettGutowski] that limits its use
to two-membranes stacks [\LambertTong,\DMPvR]. 
The discovery of the BLG model has been followed by an intense
activity concerning its interpretation and possible modifications
(either with degenerate or indefinite metric for the scalars or 
with less supersymmetry), possibly relevant for the formulation of
multiple membrane theory The literature in these directions of the
subject is large, and we refer to \eg\ ref. [\StringTalks] for a
selection of references. 

It is of course desirable to formulate a model with as much manifest
symmetry as possible. Being a maximally supersymmetric theory, the BLG
model has on-shell supersymmetry, and no finite set of auxiliary
fields. An appropriate treatment of such models --- the standard example
being $D=10$ super-Yang--Mills --- is to use pure spinors. This is the
approach that will be taken in this letter. We will in fact
investigate the most general form of the interactions, under very mild
assumptions. As expected, we recover the condition consisting of the existence
of a 3-algebra and its structure, although this derivation becomes
much easier when supersymmetry is kept manifest. It was realised early
that pure spinor techniques are relevant for supersymmetric theories
and supergravity [\NilssonPure,\HowePureI,\HowePureII]. The principles behind
and applications of pure spinors for maximally supersymmetric field
theories may be found in \eg\ refs. 
[\CederwallNilssonTsimpisI,\BerkovitsParticle,\SpinorialCohomology,\CederwallNilssonTsimpisII,\Movshev].

The first thing to decide is what the pure spinors are, and to find
the correct representations of the wave functions. The spinorial
coordinates are $\theta^{A\a}$, where $A=1,2$ is a spinor index under
the 3-dimensional Lorentz group $Spin(1,2)\approx SL(2,{\Bbb R})$ 
and $\a=1,\ldots,8$ is a
chiral spinor ${\bf 8}_s$ under the R-symmetry group $Spin(8)$. We denote the
3-dimensional vector indices $a,b,\ldots$. The superspace
torsion is $T_{A\a,B\b}^c=\g^c_{AB}\d_{\a\b}$. A pure spinor BRST
operator is generically formed as $Q=\l^{A\a}D_{A\a}$, and the purpose
of the pure spinor constraint is to make $Q$ nilpotent by projecting 
out the torsion in
$Q^2$. We see that the appropriate pure spinor constraint is
$$
(\l^{(A}\l^{B)})=0\punkt\Eqn\PureSpinorConstraint
$$
The notation for spinor contractions, denoted by parentheses, is
throughout the paper that
$Spin(8)$ indices are contracted, while $SL(2)$ indices are kept
explicit.
Very similar pure spinor constraints have been considered in ref. [\FreGrassi]. 
If we introduce Dynkin labels for representations of $sl(2)\oplus
so(8)$, $\theta$, $D$ and $\l$ transform in ${\bf 8}_s=(1)(0010)$. A
bilinear in $\l$ contains the representations
$(0)(0100)\oplus(2)(0000)\oplus(2)(0020)$, and the second of these
(the 3-dimensional vector) is
removed by the pure spinor constraint. 
The pure spinor is a (first quantised) ghost, with ghost number 1.
A pure spinor wave function is seen as a power expansion in $\l$. In
order to calculate cohomologies of $Q$ we need a list of the
representations occurring at any power $\l^n$. Using the pure spinor
constraint (\PureSpinorConstraint), one finds that these
representations are 
$$
\bigoplus_{i=0}^{[n/2]}(n-2i)(0,i,n-2i,0)\punkt\Eqn\RepsAtN
$$
This looks at first sight more complicated than the situation in \eg\
$D=10$, $N=1$ pure spinor space [\BerkovitsI], where one has one irreducible
representation at each $n$. Irreducible representations however occur
in many other situations, like $D=11$ supergravity 
[\CGNN,\CGNT,\HoweTsimpis,\SpinorialCohomology], and we will see
that the end results, the cohomologies, are quite simple. 

Let us first consider a scalar wave function. Its expansion in $\l$ 
simply contains the representations in eq. (\RepsAtN) of decreasing
ghost number $1-n$. In order to find the representations of component fields
(and field equations) one considers, as usual, the cohomology of $Q$ at zero
modes of $\partial_a$. This is a purely algebraic computation, that
can be performed by hand, but preferably by the computer-based method
of ref. [\SpinorialCohomology]. The result is given in Table 1.

\vskip2\parskip
\vbox{
$$
\vtop{\baselineskip20pt\lineskip0pt
\ialign{
$\hfill#\quad$&$\,\hfill#\hfill\,$&$\,\hfill#\hfill\,$&$\,\hfill#\hfill\,$
&$\,\hfill#\hfill\,$&$\,\hfill#\hfill\,$\cr
            &n=0    &n=1    &n=2    &n=3  &n=4 \cr
\hbox{dim}=0&(0)(0000)&\phantom{(0)(0000)}&       &          \cr
        \Fr12&\bullet&\bullet&\phantom{(0)(0000)}   \cr
           1&\bullet&(2)(0000)&\bullet&\phantom{(0)(0000)} \cr
       \Fr32&\bullet&\bullet&\bullet&\bullet&\phantom{(0)(0000)}     \cr
           2&\bullet&\bullet&(2)(0000)&\bullet&\bullet\cr
       \Fr52&\bullet&\bullet&\bullet&\bullet&\bullet\cr
           3&\bullet&\bullet&\bullet&(0)(0000)&\bullet\cr
       \Fr72&\bullet&\bullet&\bullet&\bullet&\bullet\cr
           4&\bullet&\bullet&\bullet&\bullet&\bullet\cr
}}
$$
\vskip2\parskip
\centerline{\it Table 1. The cohomology of the scalar complex.}
}
\vskip2\parskip
 
\noindent The grading $n$ is the degree of homogeneity in $\l$, 
and the vertical
direction is the expansion in the fermionic coordinates. The
superfields at different $n$ are shifted vertically 
in the table so that $Q$ acts horizontally. This cohomology describes
a Chern--Simons field, its ghost and the associated anti-fields. The
fermionic scalar wave function $\Psi$ has to be assigned ghost number 1 and
dimension 0 in order for the connection $A_a$ to have ghost number 0 and
dimension 1. 
It is essential to note that the antifield $A_a^*$ has dimension 2, so
that the equation of motion for $A$ following from $Q\Psi=0$ 
is first order in derivatives. It reads $dA=0$. There is no additional
input to the pure spinor formalism that gives the Chern--Simons
dynamics (as opposed to \eg\ Yang--Mills).
The full non-linear Chern--Simons theory follows from the non-linear
modification of the cohomology:
$$
Q\Psi+\fr2[\Psi,\Psi]=0\komma\eqn
$$
where $[\cdot,\cdot]$ is the Lie bracket.

We also need to describe the matter multiplet. It contains scalars
$\phi^I$ in
the vector representation $({\bf 1},{\bf 8}_v)=(0)(1000)$ of $Spin(8)$ 
and spinors 
$\chi^{A\dot\a}$ in the chiral
representation $({\bf 2},{\bf 8}_c)=(1)(0001)$. Since there is no gauge
symmetry, we expect the scalar fields to sit at the lowest order of
$\l$ in the wave function. We therefore try a bosonic wave function $\Phi^I$ of
dimension 1/2 and ghost number 0 in $(0)(1000)$. A pure spinor wave
function in a non-scalar representation of the structure group (such
as supergravity complexes containing the spinorial 1-form frame field,
or some non-maximally supersymmetric models [\CederwallNilssonSix])
is always subject to some further condition; the cohomology would
otherwise just be the tensor product of the wave function
representation with the cohomology of a scalar wave function. These
extra conditions typically remove ``smaller representations'', in the
same spirit as the pure spinor constraint itself. In the present case
one may postulate an additional invariance under
$$
\d_\varrho\Phi^I=(\l^A\sigma^I\varrho_A)\Eqn\RhoVariation
$$ 
for arbitrary functions 
$\varrho_A^{\dot\a}$. The effect of this further invariance (which
should be implemented the same way as the pure spinor constraint, in
the sense that
the wave function belongs to an equivalence class modulo such
functions) is that the expansion of $\Phi^I$ at order $\l^n$ (\ie,
ghost number $-n$) contains the representations
$$
\bigoplus_{i=0}^{[n/2]}(n-2i)(1,i,n-2i,0)\punkt\Eqn\PhiRepsAtN
$$
Note the similarity to the representation content of the scalar case
(\RepsAtN), the difference is only the 1 in the vector position.
This construction is vindicated by the observation that the zero-mode
cohomology is the correct one, given in Table 2. 
We observe that the
antifields have the correct dimensions and representations. The
equations of motion derived from $Q\Phi^I=0$ of course are $\square\phi^I=0$,
$(\gamma^a\partial_a\chi)^{A\dot\a}=0$.

\vskip2\parskip
\vbox{
$$
\vtop{\baselineskip20pt\lineskip0pt
\ialign{
$\hfill#\quad$&$\,\hfill#\hfill\,$&$\,\hfill#\hfill\,$&$\,\hfill#\hfill\,$
&$\,\hfill#\hfill\,$&$\,\hfill#\hfill\,$\cr
            &n=0    &n=1    &n=2    &n=3  &n=4 \cr
\hbox{dim}=\Fr12&(0)(1000)&\phantom{(0)(0000)}&       &          \cr
        1&(1)(0001)&\bullet&\phantom{(0)(0000)}   \cr
           \Fr32&\bullet&\bullet&\bullet&\phantom{(0)(0000)} \cr
       2&\bullet&(1)(0001)&\bullet&\bullet&\phantom{(0)(0000)}     \cr
           \Fr52&\bullet&(0)(1000)&\bullet&\bullet&\bullet\cr
       3&\bullet&\bullet&\bullet&\bullet&\bullet\cr
           \Fr72&\bullet&\bullet&\bullet&\bullet&\bullet\cr
}}
$$
\vskip2\parskip
\centerline{\it Table 2. The cohomology of the vector complex.}
}
\vskip2\parskip

Before introducing interactions, we would like to discuss how to write
an action. Consider first the Chern--Simons multiplet. The cohomology
contains a singlet at $\l^3\theta^3$, the position of the
antighost. This cohomology can serve as a measure with ghost number
$-3$. Strictly speaking, this is not true with the minimal set of
variables described here (analogous to the single pure spinor in
$D=10$). Unless further variables are introduced, this measure is
degenerate, since an action constructed from it will not contain
components of fields at higher powers than $\l^3$. Berkovits has shown
how to add extra variables that render the measure non-degenerate
without changing the cohomology [\BerkovitsNonMinimal]. 
We will not go further into this
procedure, but note that it is clear that the corresponding
construction works also in the present setting. Using a non-degenerate
measure in
non-minimal pure spinor space, one may write a Lagrangian for the
Chern--Simons multiplet:
$$
{{\cal L}}_{\hbox{\sixrm CS}}=\fr2\<\Psi,Q\Psi+\fr3[\Psi,\Psi]\>\komma\eqn
$$
where $\<\cdot,\cdot\>$ is a trace on the Lie algebra and
$[\cdot,\cdot]$ the Lie bracket.

How does one write a (linearised) Lagrangian for the scalar multiplet?
There are two issues, that turn out to be solved simultaneously. The
wave function $\Phi^I$ is bosonic, which 
excludes an expression like $\Phi^IQ\Phi^I$, rather one needs to
contract the $SO(8)$ vector indices with some antisymmetric tensor. In
addition, to get ghost number 3 (there is no other scalar cohomology
available), one needs an insertion of two powers of $\l$. The unique
possibility seems to be 
$$
{{\cal L}}_{\hbox{\sixrm free scalar}}=\fr2M_{IJ}\Phi^IQ\Phi^J\komma
\Eqn\ScalarLagrangian
$$ 
with $M_{IJ}=\e_{AB}(\l^A\sigma_{IJ}\l^B)$. One now has to check
that the equations of motion $M_{IJ}Q\Phi^J=0$ are equivalent to
$Q\Phi^I=0$. This happens to be true exactly thanks to the invariance
(\RhoVariation). Namely, any part of $\Phi^I$ of the form 
$(\l^A\sigma^I\varrho_A)$ drops out of $M_{IJ}\Phi^J$ due to the Fierz
identity $\e_{AB}(\l^A\sigma_{IJ}\l^B)(\sigma^J\l^C)^{\dot a}=0$,
which easily is shown to hold for pure spinors (but not general
ones). This shows that the form (\ScalarLagrangian) of the scalar field
Lagrangian is good, and gives yet another reason for the choice of the
content of the wave function implied by the equivalence classes of
eq. (\RhoVariation). 
The factor of $\l^2$ in eq. (\ScalarLagrangian), whose necessity we
have already given a number of arguments for, will turn out to be crucial in
checking the consistency of the interacting Lagrangian.
This new mechanism, with insertions of $\l$'s in the action and the
corresponding consistent modding out of representations in the wave
function, may turn out to have applications in different settings, \eg\ in the
context of supergravity. 

We now have a working supersymmetric description of the
non-interacting fields (in the Chern--Simons case self-interaction is
included). The next step is to let the scalars transform also under
some representation $\R$ of the Lie algebra of $\Psi$. We introduce traces
$\<\cdot,\cdot\>_{\xadj}$ for the adjoint and $\<\cdot,\cdot\>_{\xR}$ for
${\bf R}$, which has to be an orthogonal representation. 
The Lie bracket is written 
$[\cdot,\cdot]$ and the action of an element $T\in\adj$ on 
$x\in{\bf R}$ is denoted $T\cdot x$. We have the obvious relations
$\<x,T\cdot y\>_{\xR}=-\<T\cdot x,y\>_{\xR}$ and $T\cdot(U\cdot x)-U\cdot(T\cdot
x)=[T,U]\cdot x$. 

One term in the Lagrangian, apart from the ones already discussed,
comes from the ``covariantisation'' of $Q\Phi^I$ to
$(Q+\Psi\cdot)\Phi^I$. We then have
$$
{{\cal L}}=\<\Psi,Q\Psi+\fr3[\Psi,\Psi]\>_{\xadj}
    +\fr2M_{IJ}\<\Phi^I,Q\Phi^J+\Psi\cdot\Phi^J\>_{\xR}\punkt\Eqn\Lagrangian
$$
When trying to find other terms for an Ansatz with dimension 0 and
ghost number 3, one finds that the terms 
already present in eq. (\Lagrangian) exhaust the list of possible
ones, as long as no
dimensionful constant is introduced and no explicit fermionic
derivatives or derivatives with respect to $\l$ are allowed to enter
(neither of these are wanted, unless we search for higher derivative
modifications). Essentially, counting of dimension demands one power
of $\l$ for each $\Phi$. Unless the number of $\Phi$'s is even, one
can not form a scalar. Ghost number counting then 
limits the number of $\l$'s, and consequently of $\Phi$'s, to zero or
two. It turns out that the Lagrangian (\Lagrangian) is the full
answer, but before making that claim we have to examine its
consistency. 

All fields are obtained from cohomologies, \ie, free fields are
solutions of ``$Q(\hbox{field})=0$'' modulo gauge transformations 
``$\d(\hbox{field})=Q(\hbox{gauge parameter})$''. When interactions are
turned on, there still has to exist a gauge invariance of the form
``$\d(\hbox{field})=Q$(gauge parameter) + (interaction terms)''. 
This applies also
for the scalar wave function. Consistency of the interactions 
is proven if one finds the
invariance under such gauge transformations. This should be equivalent
to demanding that the action satisfies a Batalin--Vilkovisky master
equation (it would be interesting to find explicit generic expressions for the
anti-bracket in the pure spinor framework).

The gauge invariance corresponding to the Chern--Simons field is the
simplest part. The extra ``connection term'' in the minimal coupling
of the matter multiplet was introduced to ensure
this invariance, and the action is indeed (almost manifestly) invariant under
$$
\eqalign{
\d_\Lambda\Psi&=Q\Psi-[\Lambda,\Psi]\komma\hfill\cr
\d_\Lambda\Phi^I&=-\Lambda\cdot\Phi^I\komma\hfill\cr}
\eqn
$$
up to a $Q$-exact term (a ``total derivative'').

A general variation of the Lagrangian is
$$
\d{{\cal L}}
=\<\d\Psi,Q\Psi+\fr2[\Psi,\Psi]+\fr2M_{IJ}\{\Phi^I,\Phi^J\}\>_{\xadj}
+M_{IJ}\<\d\Phi^I,Q\Phi^J+\Psi\cdot\Phi^J\>_{\xR}\komma\eqn
$$
where we have introduced the notation $\{\cdot,\cdot\}$ for the
formation of an adjoint element from the antisymmetric product of two
elements in $\R$ via 
$\<x,T\cdot y\>_{\xR}=\<T,\{x,y\}\>_{\xadj}$.
The gauge transformation corresponding to the matter wave function is
$$
\eqalign{
\d_\Xi\Psi&=-M_{IJ}\{\Phi^I,\Xi^J\}\komma\hfill\cr
\d_\Xi\Phi^I&=Q\Xi^I+\Psi\cdot\Xi^I\punkt\hfill\cr}
\eqn
$$
Roughly speaking, the second of these equations is like a covariant
derivative. When applied to the covariantised matter kinetic term a
field strength $(Q+\Psi)^2$ arises, and this is cancelled by the
appropriate transformation of $\Psi$, whose Chern--Simons term varies
to the field strength. The only term not immediately cancelled comes
from the variation of $\Psi$ in the matter kinetic term. It
contains four powers of $\l$ and is proportional to
$$
M_{IJ}M_{KL}\<\{\Phi^I,\Phi^J\},\{\Phi^K,\Xi^L\}\>_{\xadj}\punkt\Eqn\Remainder
$$
For general choices of the representation $\R$ and of
$\<\cdot,\cdot\>_{\xadj}$ this term will not vanish, and there is no
consistent interaction. The pure spinors give information on allowed
structures. The fourth power of a pure spinor contains the
representations
$(0)(0200)\oplus(2)(0120)\oplus(4)(0040)$. This means that the product
$M_{IJ}M_{KL}$, being $Spin(1,2)$ scalar, is in $(0)(0200)$, and
as a consequence $M_{[IJ}M_{KL]}=0$. If, and only if, the expression 
$\<\{x,y\},\{z,w\}\>_{\xadj}$ is completely antisymmetric in its
arguments, the potentially problematic term given by eq. (\Remainder)
vanishes. It is then convenient to introduce the antisymmetric 3-bracket 
$\leftbr\cdot,\cdot,\cdot\rightbr$ via $\leftbr
a,b,c\rightbr=\{a,b\}\cdot c$,
or equivalently,
$\<\{x,a\},\{b,c\}\>_{\xadj}=\<x,\leftbr a,b,c\rightbr\>_{\xR}$. 

It is interesting to examine the (super-)algebra of gauge
transformations. The commutators $[\d_\Lambda,\d_{\Lambda'}]$ and  
$[\d_\Lambda,\d_{\Xi}]$ are ``covariant''. The remaining one, 
$[\d_\Xi,\d_{\Xi'}]$, requires calculation, which when acting on
$\Psi$ yields
$[\d_\Xi,\d_{\Xi'}]\Psi=\d_{\Lambda(\Xi,\Xi')}\Psi$, where 
$\Lambda(\Xi,\Xi')=-M_{IJ}\{\Xi,\Xi'\}$. The structure is like a
superalgebra, where the anticommutator of two fermionic gauge
transformations gives a bosonic one. Performing the same calculation on
$\Phi^I$ gives
$[\d_\Xi,\d_{\Xi'}]\Phi^I=-2M_{JK}\leftbr\Xi^{[K},{\Xi'}{}^{I]},\Phi^J\rightbr
=\d_{\Lambda(\Xi,\Xi')}\Phi^I-3M_{JK}\leftbr\Xi^{[I},\Xi'{}^{J},\Phi^{K]}\rightbr$.
The last term is an equivalence transformation of the type
(\RhoVariation). 
Any transformation $\d\Phi^I=s^{IJK}M_{JK}$ with antisymmetric $s^{IJK}$
is trivial, as seen by setting the parameter $\varrho\sim
s^{IJK}\sigma_{IJK}\l$. 

We have thus established that the existence of the 3-algebra is the
possibility allowed by the pure spinors for the modified (interacting)
cohomology to exist. The 3-bracket of course has to satisfy the
fundamental identity, but it provides no further information, as it
follows from the structure already defined (more precisely, from 
the fact that the the 3-bracket of three elements in $\R$
itself transforms in $\R$ under a transformation with an element in
$\adj$ defined by two elements in $\R$ via the curly bracket).
Once the structure of the pure spinor wave functions is established, 
the calculation is considerably simpler than in the component
formalism --- apart from a single term giving the
information about the need of an antisymmetric 3-bracket the
symmetries are essentially covariantly realised. A striking property
of the Lagrangian is that it only contains terms quadratic and cubic
in fields, and that it, in strong contrast to the component action, 
essentially consists of a Chern--Simons term
and minimally coupled matter. 

The component action contains a 6-point coupling of
scalars and a coupling of two scalars and two fermions, 
appropriate for a conformal model. The interactions between
component fields arise after elimination of auxiliary fields,
much in the same way as the $D=10$ super-Yang--Mills dynamics, with
4-point couplings, arises from a Chern--Simons-like action for the
pure spinor wave function. This calculation is of standard superspace
type and goes schematically as follows. Let
$\phi^I$ be the ghost number 0 part of $\Phi^I$. The matter equations
of motion then all follow from the equation ${{\cal
D}}\phi|_{(1)(1010)}=0$, where ${{\cal D}}$ includes the ghost number 0
gauge superfield. This is solved by ${{\cal D}}\phi={{\cal
D}}\phi|_{(1)(0001)}=\chi$. Acting with another fermionic covariant derivative
yields ${{\cal D}}\chi={{\cal
D}}^2\phi=(\partial_A+f)\phi=\partial_A\phi+\phi^3$, where $f$ is the
field strength with two fermionic components, which by the equation of
motion for the gauge superfield is proportional to $\phi^2$. Yet
another fermionic derivative gives the equation of motion for $\chi$,
schematically reading $\dslash_A\chi+\phi^2\chi=0$, etc. 
The formalism
also makes it completely clear why the Chern--Simons gauge field is
needed to close the supersymmetry algebra in the component
formulation: the interactions between matter fields arise from
elimination of the dimension-1 superspace component $f_{A\a,B\b}$ of the
field strength.

\acknowledgements The author would like to thank Dimitrios Tsimpis,
Ulf Gran and Bengt E.W. Nilsson for discussions.


\refout
\end